\documentclass[11pt,twocolumn]{article}

\newcommand{\ph}[1]{\medbreak \noindent {\bf #1}}

\usepackage{graphicx}

\usepackage{color}

\title{The Design of a Community Science Cloud: \protect\\
The Open Science Data Cloud Perspective}

\author{Robert L. Grossman\footnote{Robert L. Grossman is the corresponding author.} 
\and Matthew Greenway \and Allison P. Heath 
\and Ray Powell \and Rafael D. Suarez \and Walt Wells \and Kevin White \and
University of Chicago, Chicago, IL, USA
\protect\\
\protect\\
Malcolm Atkinson \and Iraklis Klampanos \protect\\
University of Edinburgh, Edinburgh, UK
\protect\\
\protect\\
Heidi L. Alvarez, Florida International University, Miami, FL, USA
\protect\\
\protect\\
Christine Harvey, Richard Stockton College of New Jersey, Galloway, NJ, USA
\protect\\
\protect\\
Joe J. Mambretti, Northwestern University, Evanston, IL, USA
}

\date{October 12, 2012}

\begin{document}
\maketitle

\begin{abstract}
In this paper we describe the design, and implementation of
the Open Science Data Cloud, or OSDC. The goal of the OSDC is to
provide petabyte-scale data cloud infrastructure and related
services for scientists working with large quantities of
data. Currently, the OSDC consists of more than 2000 cores and 2 PB
of storage distributed across four data centers connected by 10G
networks. We discuss some of the lessons learned during the past three years of operation
and describe the software stacks used in the OSDC.  We also
describe some of the research projects in
biology, the earth sciences, and social sciences enabled by the OSDC.
\end{abstract}

\ph{Keywords:}
data intensive computing, cloud computing, science clouds, open data

\section{Introduction}
The Open Science Data Cloud or OSDC (www.opensciencedatacloud.org) is
a petabyte-scale science cloud managed and operated by the Open Cloud
Consortium (OCC) that has been in operation for approximately three years.
The OCC is a not-for-profit that develops and operates cloud computing
infrastructure for the research community.  

The OSDC allows scientists to manage, analyze, share and archive their
datasets, even if they are large.  Datasets can be downloaded from the
OSDC by anyone.  Small amounts of computing infrastructure are
available without cost so that any researcher can compute over the
data managed by the OSDC.  Larger amounts of computing infrastructure
are made available to researchers at cost.  In addition, larger
amounts of computing resources are also made available to research
projects through a selection process so that interested projects can
use the OSDC to manage and analyze their data.

The OSDC is not only designed to provide a long term persistent home
for scientific data, but also to provide a platform or ``instrument''
for data intensive science so that new types of data intensive
algorithms can be developed, tested, and used over large amounts of
heterogeneous scientific data.

The OSDC is a distributed facility that spans four data centers
connected by 10G networks.  Two data centers are in Chicago, one is at
the Livermore Valley Open Campus (LVOC), and one is at the AMPATH
facility in Miami.

\section{Science Clouds}
Although NIST has provided a definition of a cloud \cite{Mell:2011}, it is probably most helpful for
the purposes here to think of a cloud as a warehouse scale computing facility \cite{Barroso:2009} that
provides services.  Usually, these services are self-provisioned.  The services may be infrastructure 
services, such as compute and storage resources, or software services, such as data analysis services.
We will use the term {\em Cloud Service Provider (CSP)} to refer to the provider of these services.

There are quite a few commercial CSPs and a growing number of CSPs focused on serving the needs
of scientists.  Table~\ref{table:sci-csp} shows some of the important differences between commercial
CSPs and those serving the needs of researchers.

\begin{table*}[ht]
\begin{center}
\begin{tabular}{|p{1.0in}|p{2.0in}|p{2.0in}|}\hline
& {\bf Commercial CSP} & {\bf Science CSP} \\ \hline
Point of view & Make a profit by providing services.  
& Democratize access to data. Support data driven discovery.  Provide long term archiving of data.  \\ \hline
Computing and storage & scale out computing and object based storage & also support data intensive computing 
and high performance storage \\ \hline
Flows & lots of small web flows & also large incoming and outgoing data flows \\ \hline
Accounting & essential & essential \\ \hline
Lock in & lock in is good & important to support moving data and computation between CSPs \\ \hline
\end{tabular}
\end{center}
\caption{Some of the differences between Commercial CSPs and Science CSP serving the research community.}
\label{table:sci-csp}
\end{table*}

\begin{figure}[ht]
\centering
\includegraphics[scale=0.4]{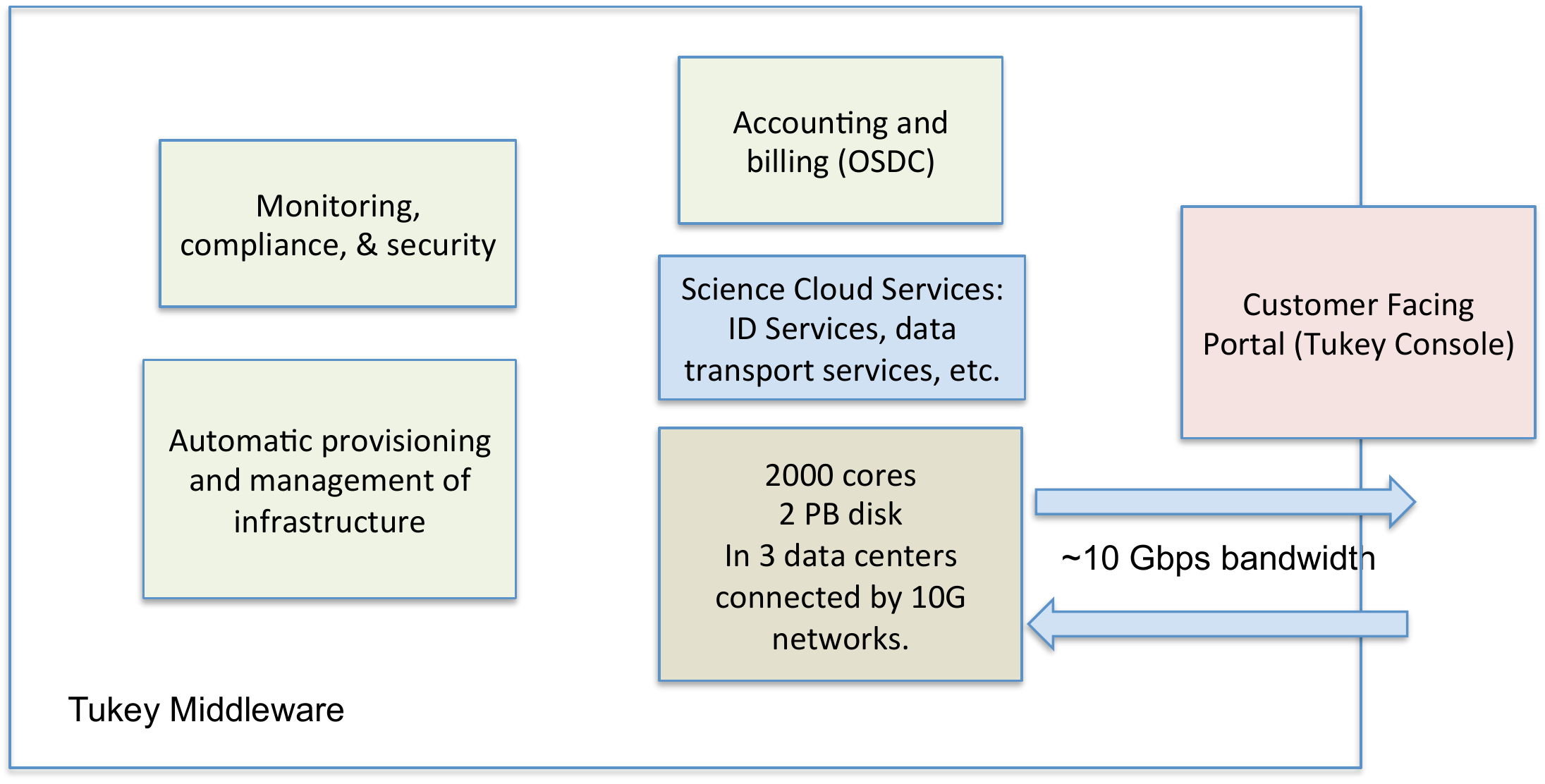}
\caption{Tukey provides the link between the users and services
  provided by the OSDC.  From the user's perspective they log into the
  Tukey Console (a web application) and have immediate access to the services
listed. This is accomplished by the web application
  interfacing with the middleware authorizing users and performing API
  translations between various cloud software stacks for virtual
  machine provisioning and determining usage.}
\label{figure:arch}
\end{figure}

\ph{CSOC.}  Essentially, the OSDC systems group operates what you
might call a {\em Cloud Services Operation Center} (CSOC) to manage
and operate the OSDC data centers, cloud services, and related
administrative services, as well as to support OSDC researchers using
the OSDC.  A CSOC is roughly analogous to a Network Operations Center
(NOC).  Much as NOCs enable the operations of the high performance
research networks that have served an important role in today's
cyberinfrastructure, CSOCs have the potential to play an analogous role 
for high performance science clouds in the future.

\begin{table*}[ht]
\begin{center}
\begin{tabular}{|p{2.0in}|p{2.0in}|p{2.0in}|} \hline
{\bf Resource} & {\bf Type} & {\bf Size} \\ \hline
OSDC-Adler \& Sullivan & OpenStack \& Eucalyptus based utility cloud & 1248 cores and 1.2PB
disk  \\ \hline
OSDC-Root & Storage cloud & approximately 1 PB of disk \\ \hline
OCC-Y & Hadoop data cloud & 928 cores and 1.0 PB disk \\ \hline
OCC-Matsu & Hadoop data cloud & approximately 120 cores and 100 TB \\ \hline
\end{tabular}
\end{center}
\caption{Summary of resources operated by the OCC.}
\label{table:occ-resources}
\end{table*}

\section{Open Science Data Cloud}

\subsection{Open Cloud Consortium}  The Open Cloud Consortium (OCC) is a
501(c)(3) not-for-profit supporting the scientific community by
operating cloud infrastructure to support scientific, medical,  health
care, and environmental research.  OCC members include universities,
such as the University of Chicago and Northwestern University;

NASA; and foreign partners, such as
AIST. Table~\ref{table:occ-resources} contains a summary of the
resources currently  operated by the OCC (we will be more than
doubling these resources in 2013).

The Open Science Data Cloud (OSDC) is one of the projects managed
by the OCC.  

\subsection{OSDC Approach}  The OSDC currently manages about a PB of data
for the researchers that it supports and operates a PB-scale Eucalyptus,
OpenStack, and Hadoop-based infrastructure so that researchers can
compute over their data.  It provides tools so that researchers can
upload and analyze datasets, integrate their data with data from
research collaborators, and integrate their data with public data for
analysis.

The approach adopted by the not-for-profit OSDC is modest: 
1) Use a community of users and data
curators to identify data to add to OSDC.  This data is managed by
the OSDC-Root cluster. Data is made available via a high
performance storage system (currently GlusterFS).  2) Use permanent
IDs to identify this data and associate metadata with these IDs.  3)
Support permissions so that colleagues can access this data
prior to its public release and to support analysis of access-controlled data.  
4) Support both file-based descriptors and
APIs to access the data.  5) Make available computing images
via infrastructure as a service that contain the software tools and
applications commonly used by a community.  6) Provide mechanisms to
both import and export data and the associated computing environment
so that researchers can easily move their computing
infrastructures between science clouds.  7) Identify a sustainable
level of investment in computing infrastructure and operations and
invest this amount each year.  8) Provide general support for a
limited number of applications.   9) Encourage OSDC users and community of users to
develop and support their own tools and applications.  

\section{Research Communities\protect\\ 
Served}
\label{section:disciplines}

The OCC's Open Science Data Cloud (OSDC) currently serves
multiple disciplines that use big data, including the earth sciences,
biological sciences,  social sciences, and digital humanities.  In
addition, the OSDC provides backup for a variety of multiple terabyte
size datasets, including the Sloan Digital Sky Survey, modENCODE,
and ENCODE.\footnote{Many of these datasets are available from
opensciencedatacloud.org/publicdata.}

\subsection{Biological Sciences}  The OSDC hosts over 400 TB for the
biological sciences community, including the 1000 Genomes dataset,
many of the datasets available from NIH's NCBI, and the Protein Data
Bank.  

The OSDC provides biologists working with big data several
benefits.   First, the OSDC make it easy to store and share
data, which is particularly beneficial for large collaborations in
which many different research groups are analyzing the same data using
different techniques. With the OSDC, different groups can analyze large
biological datasets without the necessity of each group downloading
the data.

Second, the OSDC makes available previously published data as OSDC
public datasets and also makes available OSDC managed virtual machine
images that include the analysis tools and pipelines used by the
different research groups.  In practice, that has significantly reduced
the time required for a research group to analyze or reanalyze data.
This approach is also quite helpful for those groups that do not want
to code and maintain their own pipelines.

Third, by supporting the integrated storage of both raw and analysed
data in OSDC, it has been easier for biologists to combine different
datasets for integrative analysis.

As mentioned, the OSDC has served as a back up facility for a number
of research projects.  For example, the OSDC was able to recover data
for the modENCODE after an unusual failure at their Data Coordinating
Center (DCC) and their back up site. Beginning later this year, the
OSDC will be a backup site with cloud enabled computation for the
ENCODE Project, which over four years is expected to produce over 500
TB of data.

An OSDC project called Bionimbus (www.bionimbus.org) has developed a
cloud-based infrastructure for managing, analyzing, archiving, and
sharing large genomic datases.  Bionimbus is used by a number of
projects, including modENCODE and the T2D-Genes consortia.  There are
also secure, private Bionombus clouds that are designed to hold
controlled data, such as human genomic data.

\subsection{Earth Sciences}  Project Matsu (matsu.opencloudconsortium.org) is
a joint research project with the NASA that is developing cloud based
infrastructure for processing satellite image data.  With support from
the OSDC, scientists from NASA's EO-1 mission have ported the
processing of Level 0 and Level 1 data from EO-1's ALI and Hyperion
instruments to the OSDC-Sullivan Cloud.  We are also using OSDC-Root
to archive data on a go forward basis in the OSDC.  We currently have
over three year’s worth of data from EO-1 in the OSDC (approximately
30 TB).  Project Matsu is also developing analytics for detecting fire
and floods and distributing this information to interested parties.

\begin{figure}
\centering
\includegraphics[scale=0.50]{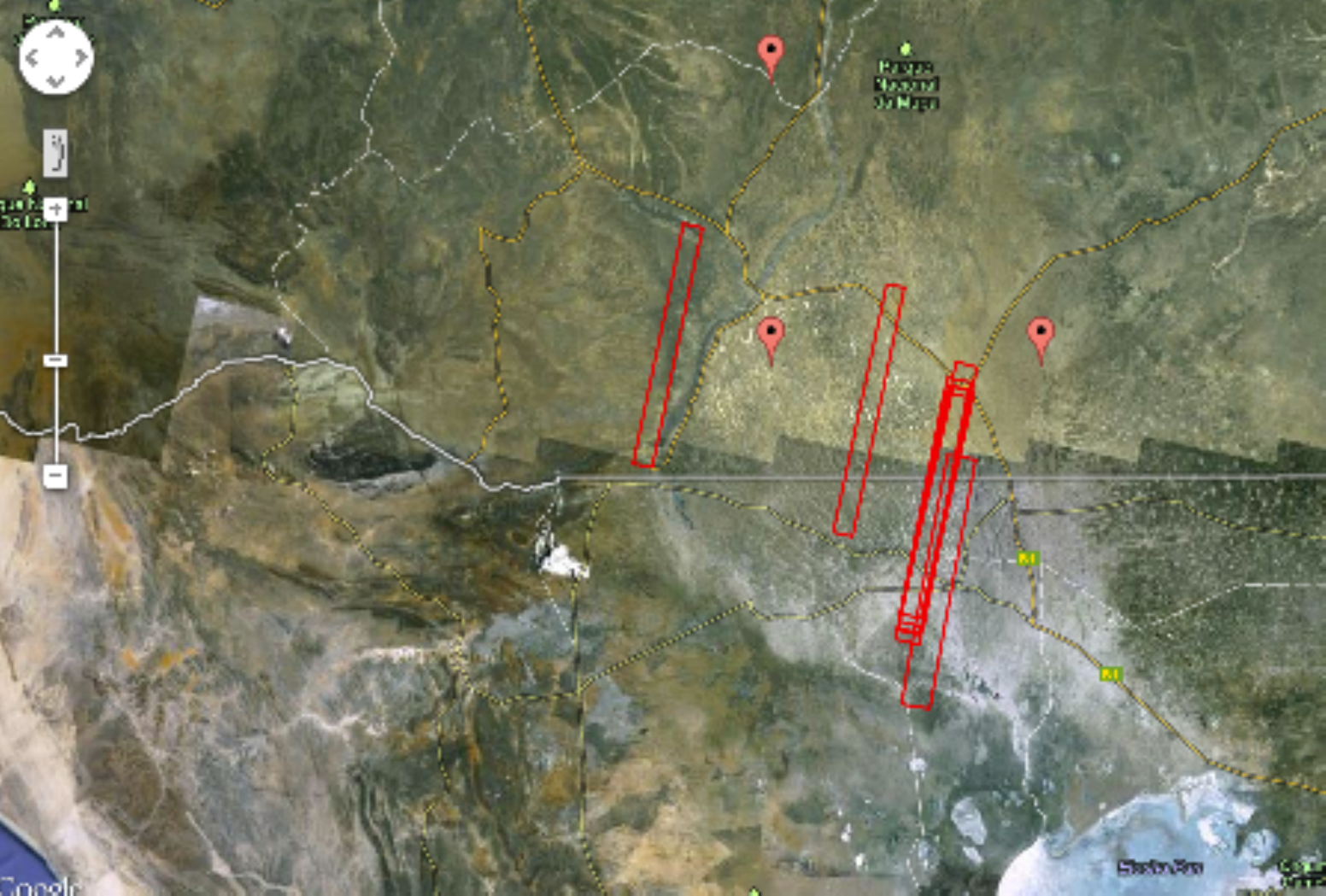}
\caption{The OSDC is used to process data from NASA's EO-1 satellite.
The map shows some of the tiles from EO-1's Hyperion instrument over
Namibia, where OSDC researchers are developing algorithms for quickly
detecting floods.}
\label{figure:matsu}
\end{figure}

\subsection{Digital humanities}  The OSDC supports Bookworm
(arxiv.culturomics.org), which is being developed by Harvard's
Cultural Observatory and offers a way to interact with digitized book
content and full text search. Bookworm uses ngrams extracted from
books in the public domain and integrates library metadata, including
genre, author information, publication place and date.

\subsection{Social Sciences}  
The OSDC hosts a variety of datasets available from the
social sciences, including the U.S. Census, Current Population Survey,
General Social Survey, and a wide variety of datasets from the
Inter-University Consortium for Political and Social Research (ICPSR),
the world’s largest repository for social science data. 

\subsection{Information sciences}  Finally, the OSDC has a variety of
datasets that are useful to researchers developing big data
algorithms, systems for processing big data, and applications for
moving big datasets.  These data sets include the Common Crawl, 
the results of focused crawls, the Enron datasets, datasets provided
by the City of Chicago, etc.

In addition, the OCC runs the OCC-Y cluster for eight computer science
departments in the U.S. that were formerly supported by the Yahoo-NSF
M45 Project, including Carnegie Mellon University and the University
of California at Berkeley.   The OCC-Y Cluster was donated by Yahoo!
to the OCC to support big data research.

\section{Tukey}
The current components of the OSDC can be roughly partitioned into two
major categories: services for users, and backend components that
enable data intensive computing.

The OSDC user services include the ability to provision virtual
machines, access usage and billing information, share files, and
access to a key service and public datasets. All of the OSDC user
services are tied together by \emph{Tukey}, an application we have
developed to provide a centralized and intuitive web interface for
accessing public and private cloud services. Figure \ref{figure:arch}
provides a high level overview of how Tukey provides OSDC user
services.

Tukey powers the OSDC web-based user interface. Tukey is composed of a
customizable web application and middleware that enables uniform
access to the cloud services by the Tukey Console web application.

\subsection{Tukey Console}
The Tukey Console is a web application based on Django and utilizes
the Tukey middleware to provide easy access to cloud services for
users. The project began as an extension of Horizon, OpenStack's
Dashboard. However, the need to support different authentication
methods and other cloud software stacks required forking from the
Horizon project. The core functionality of the web application is
virtual machine provisioning with usage and billing information. We
have also developed optional modules to provide web interfaces to
other OSDC capabilities. These include file sharing management
(Section \ref{sec:permissions}) and public data set management
(Section \ref{sec:datasets}).  Since all components of the application
communicate through sockets, they can be moved to separate servers and
use encrypted channels on our private network.

\subsection{Tukey Middleware}
The middleware portion of Tukey provides the ability to authenticate
users and interface with various cloud software stacks. It consists of
HTTP based proxies for authentication and API translations that sit
between the Tukey web application and the cloud software stacks. This
design allows it to be extensible to use other forms of authorization
and cloud software stacks. Currently, the software can handle
authentication via Shibboleth or OpenID and can interface with
OpenStack and Eucalyptus based clouds. After receiving either a
Shibboleth or OpenID identifier, the proxy looks for the cloud
credentials associated with the identifier in the user database. These
credentials are securely provided to the API translation proxies. The
translation proxies take in requests based on the OpenStack API and
then issue commands to each cloud based on mappings outlined in
configuration files for each cloud. The result of each request is then
transformed according to the rules of the configuration file, tagged
with the cloud name and aggregated into a JSON response that matches
the format of the OpenStack API.

\section{OSDC User Services}

\subsection{Dataset IDs}
As part of our efforts to provide persistent, long-term access to
scientific data we have developed a cloud service that provides IDs to
datasets based upon ARK Keys \cite{Kunze:2006}. We obtained a
registered Name Assigning Authority Number (NAAN) and have begun
assigning ARKs to the data in the OSDC. Currently, the key service can resolve persistent
identifiers and provide metadata based on ARK inflections. The service
can also be used with other persistent identifiers, which we may add in
the future.

\subsection{File Sharing and Permissions}
\label{sec:permissions}

One of the goals of the OSDC is to offer a platform where researchers
can easily collaborate on big data projects. In addition to hosting
large public datasets we also allow our users to share private
datasets with specific users or groups of their choosing. Toward this
goal, we have developed a functional prototype for distributed file
sharing, with access control based on users, groups, and
\emph{file-collection} objects. Users have the ability to create and
modify groups. A file-collection object can be a file, a collection of
files, or a collection of collections. This hierarchical structure
provides a foundation for users to manage projects and associated
datasets. In the prototype implementation, users share files by adding
them to a designated directory. This directory is monitored by a
daemon process that propagates file information to a database. Users
then utilize the OSDC web interface to grant permissions to users or
groups on file-collection objects. The system serves the files using
the WebDAV protocol while referencing the database backend. Users can
access shared files on the OSDC by mounting the WebDAV file system
with their own credentials.

\subsection{Public Datasets}
\label{sec:datasets}

The OSDC currently hosts more than 600 TB of public datasets from a
number of disciplines as discussed in Section~\ref{section:disciplines}.
One of Tukey's modules allows a data curator to manage the
dataset and the associated metadata. This information is then
published online so users can browse and search the datasets. The
datasets are stored on a GlusterFS share, described in Section
\ref{sec:storage}, so OSDC users have immediate access to all of the
public datasets. The data is freely available for download, including
over high performance networks via StarLight (www.startap.net)

\subsection{Billing and Accounting}
One of the lessons learned from early OSDC operations is that even
basic billing and accounting are effective limiting bad behavior and
providing incentives to properly share resources. We currently bill
based on core hours and storage usage. For OSDC-Adler and
OSDC-Sullivan, we poll every minute to see the number and types of
virtual machine a user has provisioned and then use this information
to calculate the core hours. Storage is
checked per user once a day. We plan to roll out similar
billing and accounting on the Hadoop clusters. Our billing cycle is
monthly and users can check their current usage via the OSDC web
interface.

\section{OSDC Backend Services}

\begin{figure*}[ht]
  \centering
    \includegraphics[width=0.7\textwidth]{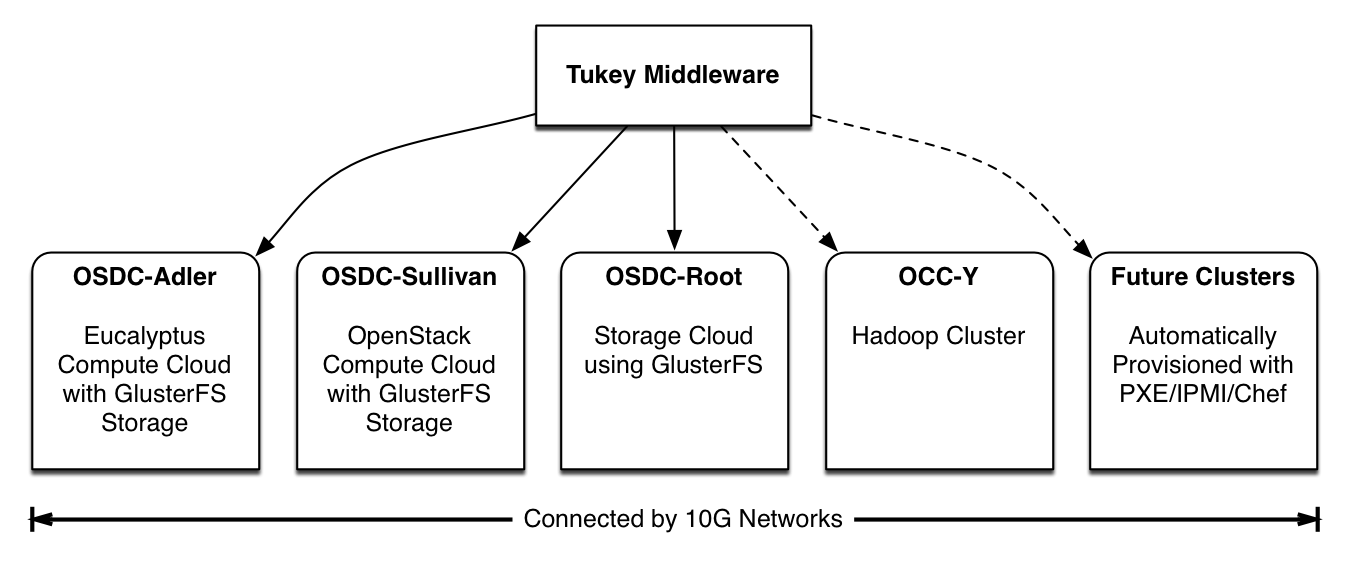}
    \caption{A diagram of the current OSDC clusters, with the solid
      arrows indicating systems fully operational and accessible with
      Tukey. The Hadoop clusters are operational and support some of
      the Tukey services but not all of them.}
  \label{fig:backend}
\end{figure*}

The OSDC is composed of compute clouds running OpenStack, Eucalyptus and
Hadoop, with GlusterFS powering the storage on OpenStack and
Eucalyptus. There are also storage racks that primarily host large
GlusterFS shares. As discussed in the previous section, Tukey's
middleware links all of these systems to the OSDC web interface,
diagrammed in Figure \ref{fig:backend}. This section discusses the
design decisions and software used to make the OSDC a platform for
scientific discovery.

\subsection{High Performance Distributed Storage}
\label{sec:storage}
A key component of the OSDC is providing file-based high performance
access to large collections of data. Despite many advances in this
realm, there are still a limited number of reliable open source file
systems that can support these requirements. Our initial experiences
with GlusterFS (version 3.1) were mixed; for example there was a bug
in mirroring that caused some data loss and forced us to stop using
mirroring. However, we now currently use version 3.3 and have observed
improvements in stability and functionality. We are using GlusterFS on
OSDC-Adler (156 TB), OSDC-Sullivan (38 TB), and OSDC-Root (459 TB) as
the primary data stores. Since
users have root access on their virtual machines we cannot allow them
to mount the GlusterFS shares directly, as the current implementation
of GlusterFS would allow them root access on the whole share. Therefore,
the GlusterFS shares are exported to the virtual machine using Samba,
which controls the permissions.

\subsection{High Speed Data Transfers}
The OSDC is a distributed facility connected by 10G networks so high
speed transport protocols are important for enabling users to import
data data, export data, and to move
data around flexibly in their analysis
processes. For example, one project generates and preprocesses their
data on OSDC-Adler Eucalyptus cluster and then sends it to the OCC-Matsu Hadoop cluster for
further analysis. Each time this is performed they have to move
several terabytes. To ease the burden of large data transfers, we have
developed UDR, a tool that provides the familiar interface of rsync
while utilizing the high performance UDT protocol.

We monitored transfer speeds of a 108 GB and a 1.1 TB dataset, between
OSDC nodes located in Chicago and LVOC with a round trip time of
104ms, using the protocols and encryption combinations shown in Table
\ref{tbl:speeds}. In addition
to measuring transfer speed, we calculate the the long distance to
local ratio (LLR). LLR provides a measure of how close the transfer
speed is of reaching the maximum throughput of the disks. The LLR is
defined as the ratio between the transfer speed and the minimum of the
source disk read speed and the target disk write speed. In our
experiments the local source disk read speed was 3072 mbit/s and local
target disk write speed was 1136 mbit/s, so the denominator for the
LLR is 1136 mbit/s. The results of these transfers using UDR and rsync (version 3.0.7) can be found in
Table \ref{tbl:speeds}. The standard encryption used by the current version of rsync is 3des. However, only blowfish encryption is currently implemented in UDR, so results for both rsync using both 3des and blowfish are included. The results show that UDR achieves 87\% and 41\% faster speeds in the unencrypted and encrypted cases, respectively, than standard rsync while retaining the same familiar interface.

\begin{table*}[!ht]\renewcommand{\arraystretch}{1.1}
\centering
\begin{tabular}{l | c c c c |}
  \cline{2-5}
  & \multicolumn{2}{|c|}{108 GB Data Set} & \multicolumn{2}{c|}{1.1 TB Data Set} \\ 
  \cline{2-5}
  & mbit/s & LLR & mbit/s & LLR\\ \hline
  \multicolumn{1}{|l|}{UDR (no encryption)} & 752 & 0.66 & 738 & 0.64\\
  \multicolumn{1}{|l|}{rsync (no encryption)} & 401 & 0.35 & 405 & 0.36\\ \hline
  \multicolumn{1}{|l|}{UDR (blowfish)} & 394 & 0.35 & 396 & 0.35 \\ 
  \multicolumn{1}{|l|}{rsync (blowfish)} & 280 & 0.25 & 281 & 0.25\\ \hline
  \multicolumn{1}{|l|}{rsync (3des)} & 284 & 0.25 & 285 & 0.25\\ \hline
\end{tabular}
\caption{Overall transfer speeds in mbit/s and the long distance to local ratio (LLR) between OSDC nodes in Chicago, IL and Livermore, CA, round trip time of 104 ms.}
\label{tbl:speeds}
\end{table*}

\subsection{Automated Provisioning}
Our first full rack installation of OpenStack was performed manually
and took over a week to complete. As the OSDC scales, this would be
unsustainable and limit our capabilities. To address this, we are
using Chef (www.opscode.com/chef), along with PXE booting and IPMI, to
fully automate provisioning with the goal of taking a full rack from
bare metal to a compute or storage cloud in much less than a day. Our
system starts with one PXE boot server, a Chef server, and a set of
servers with IPMI configured. IPMI is triggered to boot the servers,
which then pull a start-up image and boot options from the PXE boot
server. The options define the location of a preseed file which the
start-up sequence loads and uses as a guide to automate the
installation of the generic Ubuntu Server. The installation uses the
PXE boot server or a preconfigured proxy to install Ubuntu Server
directly from the repositories to avoid any updates being required
post-install. Then the installer runs a script specified at the end of
the preseed file which sets up networking on the freshly installed
system and adds another script to be run at boot. Upon rebooting, the
next script double-checks the IPMI configuration, finishes
partitioning the disk and sets up additional RAIDs as necessary,
before downloading and installing the Chef client. The Chef client
then checks in with the Chef server and runs the ``recipes'' listed
for either a management node or a compute node. As a last step, a
final clean up script runs to deliver us a fully functional OpenStack
rack.

\subsection{Monitoring Services}
Once the systems are up and running, we perform two types of
monitoring to automatically identify issues, provide alerts, and
produce reports on the status and health of the systems. The first
type of monitoring is cloud usage, such as how many instances each
user is running. We have developed an in-house application for this
purpose. The high level summary of the cloud status is made public on
the OSDC website. The second type of monitoring is system and network
status, for which we use the open source Nagios application
(www.nagios.org).

The Nagios console is browser based and Nagios uses an agent, NRPE, to monitor
the remote hosts in our environments and the services we wish to
monitor on the remote hosts.  The agent listens via TCP and
communicates back to the master server after running checks on the
remote servers that we configure using Nagios' text based
configuration files.  The master server, via the agent, asks for
checks to be run and returns the values to the master server using
binary plugins with arguments that designate the thresholds for
``Warning'' and ``Critical'' alerts.  When those thresholds are crossed,
Nagios sends alerts to the system administrators.

\section{Sustainability Model}

The OSDC is governed by a set of bylaws that are available on the web.   
Activities are organized into working groups.  For example, the OSDC Working Group
manages the Open Science Data Cloud.  There is also an OCC Board of Directors
and an OCC Advisory Committee.  

The basic philosophy of the OSDC Working group is summarized by the
following five rules:
\begin{enumerate}
\item Provide some services without charge to any interested researcher.
\item For larger groups and activities that require more OSDC
  resources, charge for these resources on a cost recovery basis.
\item Partner with university partners to gain research funding to
  tackle new projects and to develop new technology in order to
  further the mission of the OSDC.
\item Raise funding from donors and not-for-profits in order to
  provide more resources to more researchers.
\item Work to automate the operation of the OSDC as much as possible
in order to reduce the costs of operations.
\end{enumerate}

\section{Commercial Clouds}

Amazon Web Services (aws.amazon.com) provide a
large and sophisticated set of infrastructure as a service (IaaS)
capabilities, including computing (EC2) and storage (S3). The elastic
(on-demand) nature of these services has made them popular for both
business and scientific applications.  Many applications have found
that the eventual consistency model \cite{DeCandia:2007} that Amazon's
SimpleDB supports to be adequate for their application requirements.
Other providers such as Microsoft Azure and the Rackspace Cloud
provide similar capabilities.

The OSDC is designed to interoperate with commercial clouds, such as
provided by Amazon.   In general, OSDC machine images can also run on
AWS.  The OSDC also supports high performance export of data from AWS
with an in-house designed tool.

\subsection{Why not just use Amazon?}  A common question is whether
researchers should just use public clouds, such as AWS, or in addition
develop their own private and community cloud infrastructure.  After
three years of operation, it is clear that community science clouds
such as the Open Science Data Cloud are complementary to public
clouds such as AWS for the following reasons: First, science clouds
such as the OSDC are designed to work with big data.  They connect to
{\em high performance 10G and 100G networks}, they support {\em high
  performance storage}, and they have support staff that are familiar
with the problems arising when managing and transporting large
datasets.

Second, for high end users, the OSDC is
less expensive than using AWS, and for users with less demands, it is
no more expensive.  As a rough rule of thumb, when we operate an OSDC
rack\footnote{Currently, a OSDC rack contains 39 servers, each with 8
  cores and 8 TB of disk.} at approximately 80\% efficiency or greater, it is less
expensive than using Amazon for the same services.  We are not arguing
that {\em every} individual researcher should have their own cloud, but instead
that it is less expensive for a community of researchers to have
several well run community clouds that can interoperate with large
scale commercial clouds, such as run by Amazon.

Third, many research products, including data products, are just
too valuable to be entrusted exclusively to a commercial entity,
which may over time, be bought, decide to close down a line of
business, or go bankrupt.  Not-for-profit community science clouds are
a valuable complement to commercial clouds.

\section{Conclusion}
The OSDC provides research communities a high performance cloud infrastructure for data intensive science. Our design and implementation of the OSDC has been driven by the goal of providing both high performance and ease of use. The result is a number of projects utilizing the OSDC in biological sciences, earth science and digital humanities. Moving forward, we plan to expand both the infrastructure and services provided to meet the increasing data demands of scientific research. This expansion will be made possible by following our sustainability model. Additionally, we will continue to reevaluate the design and update the implementation of the OSDC based on user feedback, new technologies, lessons learned, and new data driven research projects.

\section*{Paper Status}
This is an expanded version of an early draft of the paper: Heidi L. Alvarez, Malcolm Atkinson, Robert L. Grossman, Matthew Greenway, Christine Harvey, Allison P. Heath, Iraklis Klampanos, Joe J. Mambretti, Ray Powell, Rafael D. Suarez, Walt Wells and Kevin White, The Design of a Community Science Cloud: The Open Science Data Cloud Perspective, 2012 SC Companion: High Performance Computing, Networking Storage and Analysis, ACM Press, 2012. 

\section*{Acknowledgments}
This work was supported in part by grants from Gordon and Betty Moore
Foundation, the National Science Foundation (Grant OISE - 1129076 and
CISE 1127316), and the National Institutes of Health (Grants NIGMS/NIH
P50GM081892-03A1 NIH and NIMH/NIH P50MH094267).  This work also used
the Open Cloud Consortium's OCC-Y Cluster, which was donated by Yahoo!
Inc.

\bibliographystyle{IEEEtran}
\bibliography{bib-dc}

\begin{thebibliography}{1}
\providecommand{\url}[1]{#1}
\csname url@samestyle\endcsname
\providecommand{\newblock}{\relax}
\providecommand{\bibinfo}[2]{#2}
\providecommand{\BIBentrySTDinterwordspacing}{\spaceskip=0pt\relax}
\providecommand{\BIBentryALTinterwordstretchfactor}{4}
\providecommand{\BIBentryALTinterwordspacing}{\spaceskip=\fontdimen2\font plus
\BIBentryALTinterwordstretchfactor\fontdimen3\font minus
  \fontdimen4\font\relax}
\providecommand{\BIBforeignlanguage}[2]{{%
\expandafter\ifx\csname l@#1\endcsname\relax
\typeout{** WARNING: IEEEtran.bst: No hyphenation pattern has been}%
\typeout{** loaded for the language `#1'. Using the pattern for}%
\typeout{** the default language instead.}%
\else
\language=\csname l@#1\endcsname
\fi
#2}}
\providecommand{\BIBdecl}{\relax}
\BIBdecl

\bibitem{Mell:2011}
P.~Mell and T.~Grance, ``The {NIST} definition of cloud computing, special
  publication 800-145,''
  http://csrc.nist.gov/publications/nistpubs/800-145/SP800-145.pdf, September,
  2011.

\bibitem{Barroso:2009}
L.~A. Barroso and U.~Holzle, \emph{The Datacenter as a Computer -- an
  introduction to the design of warehouse-scale machines}.\hskip 1em plus 0.5em
  minus 0.4em\relax Morgan \& Claypool Publishers, 2009.

\bibitem{Kunze:2006}
J.~Kunze and R.~Rodgers, ``The {ARK} identifier scheme,''
  www.ietf.org/internet-drafts/draft-kunze-ark-15.txt.

\bibitem{DeCandia:2007}
\BIBentryALTinterwordspacing
G.~DeCandia, D.~Hastorun, M.~Jampani, G.~Kakulapati, A.~Lakshman, A.~Pilchin,
  S.~Sivasubramanian, P.~Vosshall, and W.~Vogels, ``Dynamo: amazon's highly
  available key-value store,'' in \emph{Proceedings of twenty-first ACM SIGOPS
  symposium on Operating systems principles}, ser. SOSP '07.\hskip 1em plus
  0.5em minus 0.4em\relax New York, NY, USA: ACM, 2007, pp. 205--220. [Online].
  Available: \url{http://doi.acm.org/10.1145/1294261.1294281}
\BIBentrySTDinterwordspacing

\end{thebibliography}

\end{document}